\documentstyle[epsfig,12pt,preprint,tighten,aps]{revtex}
\begin{document}

\draft

\begin{titlepage}
\rightline{August 1999}
\centerline{to appear in Phys.Lett.B}
%\rightline{UM-P-99/xx}
\vskip 2cm
\centerline{\large \bf  
Have mirror planets been observed?}
\vskip 1.1cm
\centerline{R. Foot\footnote{Email address:
Foot@physics.unimelb.edu.au}}
\vskip .7cm
\centerline{{\it Research Centre for High Energy Physics}}
\centerline{{\it School of Physics}}
\centerline{{\it University of Melbourne}}
\centerline{{\it Parkville 3052 Australia}}
\vskip 2cm

\centerline{Abstract}
\vskip 1cm
\noindent
Over the last few years, several close orbiting ($\sim 0.05$ AU)
large mass planets ($M \sim M_{Jupiter}$)
of nearby stars have been discovered. 
Their existence has been inferred from
tiny doppler shifts in the light from the star.
We suggest that these planets may be made of mirror matter.
We also suggest that some stars such as our sun
may also have a similar amount of mirror matter,
which has escaped detection.

\end{titlepage}
\noindent
Over the last few years a number of planets orbiting nearby 
stars have been discovered (for a review and references see\cite{web}). 
Their existence has been inferred from
tiny doppler shifts in the light from the star due to its orbit
around the center of mass. The periodicity and magnitude of the doppler
shifts can be used to determine the mass and orbit of the planet.

Surprisingly about half a dozen large planets (ie. with Jupiter
sized masses or larger) have been found which have 
orbits very close to the star (between $0.04$ to $0.06$ AU). 
We summarize the data in Table 1 which lists all of the confirmed
planets with orbits less than $0.1$ AU.

\vskip 1.6cm

{\begin{center}
\begin{tabular}{|l|l|l|l|l|}
\hline
{\em STAR}$\;\;\;\;\;\;\;$
&{\em MASS (J)}$\;\;\;\;\;\;\;\;\;\;\;\;$
&{\em SEM-MAJ AXIS (AU)}$\;\;\;$ 
&{\em PERIOD (d)}$\;\;\;\;$ 
&{\em ECC}$\;\;\;\;\;\;\;$ \\
\hline
HD75289&0.42&0.046&3.51 &0.054\\
28.94 pc& & & &\\
\hline
51 Peg&0.47&0.05&4.23 &0.0\\
15.36 pc& & & &\\
\hline
HD187123&0.52&0.042&3.097 &0.03\\
49.92 pc& & & &\\
\hline
Ups And&0.71&0.059&4.617 &0.034\\
13.47 pc& & & & \\
\hline
HD217107&1.28&0.04&7.11 &0.14\\
19.72 pc& & & & \\
\hline
Tau Boo&3.87&0.0462&3.3128 &0.018\\
15.60 pc& & & & \\
\hline
HD98230&37&$\sim$ 0.06&3.98 &0.00\\
& & & & \\
\hline
HD283750&50&$\sim $ 0.04&1.79 &0.02\\
& & & & \\
\hline
\end{tabular}\end{center}}

Table Caption:
Table obtained from Ref.\cite{web} showing the mass (in units
of Jupiter mass), Semi-Major axis, period (in days) and 
eccentricity of all of the confirmed
planets with orbits less than $0.1$ AU.

\vskip 1cm

Whilst the doppler shift detection technique
is quite sensitive to large mass planets with close orbits, the fact
that any such planets with these properties exist at all is unexpected.
To date none of these nearby planets has been seen directly,
however this should be possible in the future provided that
they are made of ordinary matter.
We believe that an interesting alternative possibility exists
and that is that these close orbiting large planets might 
be made of mirror matter.
[This has also been suggested
independently by Volkas\cite{volk}. ].

The existence of mirror matter is 
well motivated from a particle physics point of view, since
these particles are predicted to exist
if parity and indeed time reversal are unbroken 
symmetries of nature\cite{ly,flv}.  The idea is 
that for each ordinary particle, such as the photon, electron, proton
and neutron, there is a corresponding mirror particle, of 
exactly the same mass as the ordinary particle. For example,
the mirror proton and the ordinary proton 
have exactly the same mass\footnote{
The mass degeneracy of ordinary and mirror matter
is only valid provided that the parity symmetry
is unbroken, which is the simplest and theoretically most
attractive possibility. For some other
possibilities, which invoke a mirror
sector where parity is broken spontaneously 
(rather than being unbroken), see Ref.\cite{other}.}.
Furthermore the mirror proton is stable for
the same reason that the ordinary proton
is stable, and that is, the interactions of the mirror
particles conserve a mirror baryon number.
The mirror particles are not produced
in Laboratory experiments just because they couple very
weakly to the ordinary particles. In the modern language of gauge
theories, the mirror particles are all singlets under 
the standard $G \equiv SU(3)\otimes SU(2)_L \otimes U(1)_Y$
gauge interactions. Instead the mirror
particles interact with a set of mirror gauge particles,
so that the gauge symmetry of the theory is doubled,
i.e. $G \otimes G$ (the ordinary particles are, of 
course, singlets under the mirror gauge symmetry)\cite{flv}.
Parity is conserved because the mirror particles experience
$V+A$ mirror weak interactions
and the ordinary particles experience the usual $V-A$ weak
interactions.  Ordinary and mirror
particles interact with each other predominately by
gravity only.

At the present time there is a range of experimental evidence
supporting the existence of mirror matter.
Firstly, it provides a natural candidate for dark matter,
which might be mirror stars (and mirror dust, planets etc)\cite{blin}. 
There is an interesting possibility
that these mirror stars have already been detected experimentally
in the MACHO experiments\cite{ii}. 
Secondly, ordinary and mirror neutrinos
are maximally mixed with each other if neutrinos have mass\cite{P}.
This nicely explains the solar and atmospheric neutrino
anomalies. The idea is also compatible with the LSND experiment\cite{P}.
Interestingly, maximal ordinary - mirror neutrino oscillations 
do not pose any problems for big bang nucleosynthesis (BBN)
and can even fit the inferred primordial abundances better
than the standard model\cite{bb}.

Of course due to the nature of mirror matter, its existence
is difficult to rigorously prove (or disprove). 
If many nearby stars have close orbiting mirror planets
then this should help establish the existence of mirror matter.
Mirror planets cannot be seen directly because they cannot
reflect the light from the star.
This is a definite and in fact testable prediction of our mirror
planet hypothesis.
Another implication of the mirror planet hypothesis
is that it may not be possible to detect the change of brightness
when the planet occults the star
(i.e. when the planet passes between the 
star and its line of sight as seen from the earth).
This is just because mirror matter may  
be completely transparent to ordinary light!
\footnote{
Actually it is possible that there is a small
electromagnetic coupling between ordinary and mirror
matter arising from photon - mirror photon kinetic mixing
(see latter discussion).
This small interaction can potentially make the mirror
planet opaque (this will obviously depend on the strength
of this interaction, i.e. the parameter $\zeta$ in Eq.(\ref{ek}) below,
on the frequency of light and the amount and chemical
composition of the mirror planet).
Also, it should be noted that
the mirror planet might be only partially opaque
leading to absorption lines. This could provide a clear
signal that the planet is made of mirror matter.
Amusingly, if this could be observed, then
it might be possible to determine
the chemical composition of the mirror planet.}.
Another implication of the mirror planet
hypothesis is that 
the orbital plane of the mirror planets may be in
a completely different plane to ordinary planets.
For example, the star Upsilon Andromedae has three
confirmed Jovian planets. One with a close orbit (listed
in the fourth row of the table) and two more distant planets with
orbits 0.83 AU and 2.5 AU\cite{web}.
If the two distant planets are made of ordinary matter than
it is likely that the these two planets should orbit
in a different plane to the close orbit planet if this
is a mirror planet.
This is just because ordinary and mirror matter 
interacts with each other predominately by gravity only.

If these close orbiting planets are made of mirror matter,
then a number of questions naturally arise.
Firstly there are arguments which suggest that
ordinary and mirror matter should be segregated
on relatively large scales. A scale of $10^5$ stars was
estimated in Ref.\cite{blin2}. This was assuming that ordinary
and mirror matter interact only gravitationally (along with
a number of other assumptions which may not always be valid).
However, it was pointed out in Ref.\cite{bob,gl,flv} that ordinary and
mirror matter can interact weakly due to photon kinetic mixing.
In field theory
this is described by the interaction Lagrangian density
\begin{equation}
{\cal L} = \zeta F^{\mu \nu} F'_{\mu \nu},
\label{ek}
\end{equation}
where $F^{\mu \nu}$ ($F'_{\mu \nu}$) is the field strength 
tensor for electromagnetism (mirror electromagnetism).
This type of Lagrangian term is gauge invariant 
and renormalizable and can exist at tree level\cite{fh,flv}
or maybe induced radiatively in models without $U(1)$ 
gauge symmetries (such as grand unified theories)\cite{bob,gl,cf}.
The effect of ordinary photon - mirror photon kinetic mixing
is to give the mirror charged particles a small electric
charge\cite{bob,gl,flv}. That is, they couple to ordinary photons with
charge $\zeta e$.
This small non-gravitational force will allow some ordinary matter
- mirror matter collisions which can dissipate energy and
help ordinary stars attract a significant (`significant' 
means of the order of 0.1 percent by mass)
amount of mirror matter during their formation.

Of course one may wonder why
some stars have mirror companions and not other stars. 
In particular their is obviously no such large mirror planet
orbiting our sun.
One possibility is that all stars in our region of the galaxy
attracted a significant amount of mirror matter during their
formation. However 
for some of these stars (including our sun) it might be 
that the mirror matter 
was so close to the ordinary matter that it
was either destroyed by tidal forces or the tidal forces 
prevented it from forming in the first place. Indeed, a mirror (or
ordinary) planet would be destroyed by tidal forces
when its radius is within the Roche limit, given by\cite{book}:
\begin{equation}
r < f_R\left({\bar \rho_s \over \bar \rho_p}\right)^{1/3}R_s,
\end{equation}
where $\bar \rho_s, \ \bar \rho_p$ are the average densities of the
star and planet respectively and $R_s$ is the radius of the star.
Also $f_R \simeq 2.5$ \cite{book}.
If the mirror planet is close enough to break apart by the
tidal forces then one may be left with rings of mirror matter
surrounding the star. This might be similar
to the rings of saturn which may have formed from the tidal break 
up of a moon or moons.
Naturally we would expect much of the mirror ring material
to have migrated to the
center of the sun by either gravitational dissipation\cite{fey},
collisions of the orbiting mirror particles with themselves
and/or through the small possible non-gravitational force 
arizing form photon-mirror photon kinetic mixing.
The latter effect can easily be estimated and any mirror
matter within the radius of the star would have migrated
to the center provided that
$\zeta \stackrel{>}{\sim} 10^{-15}$.
This is consistent with the experimental and BBN bounds
on $\zeta$\cite{gl,glc,sg} which imply that
$\zeta \stackrel{<}{\sim} 10^{-6} - 10^{-8}$.

In summary,
we have suggested that the close orbiting planets discovered in
nearby stars can be plausibly explained in terms of mirror
matter. Stars (such as our sun) without close orbit
mirror planets may also have a significant amount
of mirror matter which has broken up under
the tidal forces. In this case, most of this mirror
matter would be expected to have migrated
to the center of the sun, although some mirror material
may also exist in the form of mirror
rings (similar to the rings of saturn, except
they are made of mirror matter). 
This mirror planet hypothesis will be testable
in future experiments.

\vskip 0.4cm
\noindent
{\bf Acknowledgement}
\vskip 0.4cm
\noindent
The author thanks R. R. Volkas for his comments on the
paper and A. Ignative, G. Joshi, B. Morgan and M. Drinkwater
for discussions.
The author also thanks C. Feynman and Z. Silagadze for constructive
correspondence which have lead to significant improvements to the paper.
The author is an Australian Research Fellow.

\end{document}